\documentclass[aps, prd, amsmath, floats, floatfix, twocolumn, nofootinbib, footnote, superscriptaddress, 
]{revtex4}
\usepackage{graphicx}
\usepackage{color}
\usepackage{latexsym}

\usepackage{amssymb}

\usepackage{comment}

\newcommand{\be}{\begin{eqnarray}}
\newcommand{\ee}{\end{eqnarray}}

\begin{document}

\title{Causality in Nonlocal Gravity}

\author{Stefano Giaccari}
\email{stefanog@hit.ac.il} 
\affiliation{Department of Sciences,
Holon Institute of Technology (HIT),
52 Golomb St., Holon 5810201, Israel} 

\author{Leonardo Modesto}
\email{lmodesto@sustc.edu.cn} 
\affiliation{Department of Physics, Southern University of Science and Technology, Shenzhen 518055, China}

\date{\today}

\begin{abstract}
We study the causal structure of a class of weakly nonlocal gravitational theories (eventually coupled to matter) that are compatible with perturbative unitarity and finiteness at quantum level. In particular, we show that in nonlocal quantum gravity a Shapiro's time advance never occurs. %
Moreover, we provide a recipe to construct a general ultraviolet complete gravitational theory coupled to matter (with or without supersymmetry) compatible with causality. Therefore, nonlocal gravity is consistent with causality, as well as string theory.
\end{abstract}

\maketitle



{\it Introduction ---}
In order to solve the renormalizabilty problem of Einstein's quantum gravity a new class of weakly nonlocal gravitational theories has been proposed and extensively studied. It turns out that these theories are perturbatively unitary and finite at quantum level. Technically, the Cutkovsky rules for a unitary theory are satisfied at any perturbative order, while all the loop amplitudes are finite. 

In this paper, we follow the analysis developed by Camanho, Edelstein, Maldacena, and  Zhiboedov \cite{CEMZ} to infer about the causal structure of the theory, namely, we prove that the weakly nonlocal modifications of the three-point functions present in the nonlocal gravitational theories do not lead to causality violations. Our analysis applies to a high energy scattering process in which gravity is weakly coupled. Namely, we are in the regime 
\be
\ell_{\rm P} \ll b \ll \ell_\Lambda \, , 
\ee
where $b$ (impact parameter) is the magnitude of the impact vector $\vec{b}$, $\ell_{\rm P}$ the Planck scale, and $l_{\rm \Lambda}$ the non locality scale.

It turns out that we do not need to introduce an infinite tower of massive higher spin fields to fix the causality issue because either the local or nonlocal higher derivative operators do not affect the scattering amplitudes (which are the same of the local two-derivative theory.)

Going to the crux of the problem, the analysis developed in \cite{CEMZ} is based on the computation of Shapiro's time delay, which is one of the classical tests of general relativity (GR). We know from GR that light propagating near a compact object would suffer a time delay compared with the same propagation in flat spacetime. Therefore, if we get a negative time delay, or actually a time advancement, we have a causality violation. In \cite{CEMZ} it has been proved that the most general higher derivative theory giving rise to a causality violation is at most cubic in the Riemann tensor. The nonlocal theories can be classified according to the presence or not of such operators. Indeed, we can avoid terms affecting causality and at most cubic in the Riemann tensor, but still define a good quantum gravity theory. However, the minimal finite unitary theory shows up operators dangerous for causality when the theory is expanded 
in the number of derivatives in agreement with the analysis in \cite{CEMZ}. 

{\it The Theory ---} 
Let us remind here the class of gravitational theories we want to investigate in this paper. The action we want to study consists on four operators linear, quadratic, and quartic in the Riemann tensor, while an analytic function of the d'Alembertian operator determines the non-locality. The action reads \cite{kuzmin,tomboulis,modesto,modestoLeslaw}: 
\be
S_{\rm g}  = \frac{2}{\kappa^{2}_D}  \int  d^D x \sqrt{-g} \left[ R + G_{\mu\nu} \gamma(\Box) R^{\mu\nu} + V(\mathcal{R}) \right]  .
\label{theory}
\ee
where $\kappa_D^2 = 32 \pi G$, the form factor $\gamma(\Box)$ depends on the non-locality scale $\sigma \equiv \ell_\Lambda^2$ ($[\sigma] = -2$) and is defined by 
\be
\gamma(\Box) = \frac{e^{H(\sigma  \Box)} -1}{\Box} \, .
\ee
Moreover, the potential $V(\mathcal{R})$ is at least cubic in the curvature, but quadratic in the Ricci tensor. Here $\mathcal{R}$ is the generalized curvature and stands for $R$, Ricci or Riemann tensor, and derivatives of them properly contracted. Two examples of suitable form factors $\exp H(z)$ asymptotically polynomial in a conical region $C$ around the real axis as required by the locality of the counterterms are:
\be
&& \hspace{-0.8cm}
 H_{\rm K}(z) = \alpha \left[ \log (z )+\Gamma (0,z)+\gamma_E \right]  \, , 
 \,\,\,   {\rm Re} \, z > 0 ,     \label{Hkuzmin}
 \ee
 where $z=  -\sigma \Box$, and 
 \be
 &&\hspace{-0.8cm}
H_{\rm T}(p) = \frac{1}{2} \left[ \log \left(p^2\right)+\Gamma \left(0,p^2\right)+\gamma_E \right] \, , \,\,\,  {\rm Re} \, p^2 > 0 \, ,
\label{HTombo}
\ee
where $p\equiv p(z)$ is a polynomial of $z= - \sigma \Box$ of degree $\gamma+1$ with $\gamma > D/2$. 
The front coefficient $\alpha$ is a positive integer whose minimum value is fixed on the basis of super-renormalizability, if the form factor (\ref{Hkuzmin}) is selected. 
The form factor (\ref{Hkuzmin}) is only defined in Euclidean space, while (\ref{HTombo}) can be defined also in Minkowski space depending on $\gamma$ and on the choice of the polynomial $p(z)$. 
For the theory with form factor (\ref{HTombo}) we do not have the causality violation described in \cite{CEMZ} as it is evident expanding the form factor in Taylor series. 
Indeed, the first correction to the Einstein-Hilbert (EH) action is $\mathcal{R} \Box^n \mathcal{R}$, where $n\geqslant1$ is a positive integer. Therefore, they do not lead to causality violation as proven in \cite{CEMZ}.
Nevertheless, we will study also nonlocal theories with form factor (\ref{HTombo}). 

About the main properties of the action (\ref{theory}), the Cutkovsky rules for a unitary theory are satisfied at any perturbative order.
In even dimension we can achieve finiteness with a suitable minimal choice of the potential 
$V(\mathcal{R})$, but in odd dimension the theory is finite without need of any potential. 

In this paper we will consider also other exponential form factors, namely
$e^{- \sigma \Box}$ and $e^{ \sigma^2 \Box^2}$. These form factors do not lead to well-defined quantum theories (at least on the basis of standard perturbative quantum field theory), but they capture the main properties of the theory to understand the role of non-locality in preserving causality. Moreover, they enable us to derive analytic results.




{\em Shapiro's time delay ---} In order to evaluate the Shapiro's time delay (or advance, depending on whether we have or not a causality violation) we need the tree-level scattering amplitude/s in the eikonal limit. The latter consists on summing all the ladder (and cross ladder) diagrams in the Regge limit of large $s$,
but $t/s\ll1$ (here $s,t,u$ are the Mandelstam variables). Indeed, the lowest order approximation to the scattering amplitude is the Born approximation or tree-level approximation, which corresponds to a large impact parameter $b$ or small deviation angles.  
When we decrease the impact parameter 
other diagrams involving exchange of more gravitons turns out to be important, but if $b$ is not too small such diagrams can be summed to give the eikonal approximation \cite{Erice, GiddingsPorto}. 
In the eikonal regime, $b$ is taken small, but it is still much larger then the gravitational radius $r_s = 2 G \sqrt{s}$.
Decreasing even more $b$ we reach the Planck scale and we probe the trans Plankian regime. 
At the Planck scale in general we need non perturbative resummation techniques, but the super-renormalizable theory is asymptotically-free, while the finite theory is conformal invariant.
Therefore, the theory turns out to be perturbative or in the conformal phase at extremely high energy regime. However, when $b\leqslant r_s$ we expect the collision of particles to produce trapped surfaces, signaling the presence of black holes. Nevertheless, the analysis of this regime is beyond the scope if this paper. 

We here probe the non-locality scale, namely $b\ll \ell_{\Lambda}$, but we stay far from the Planck scale and the gravitational radius $r_s$. 

In order to evaluate the scattering amplitude in the eikonal approximation we fist need 
the amplitude in the Regge limit $t/s<<1$ (but large $s$), namely $A(s,t)$. Therefore, in the eikonal approximation the four-particles scattering amplitude is exponential in the impact parameter space and the result is \cite{Kabat,Ciafaloni}: 
\be
i A_{\rm eik} = 2 s \int d^{D-2} \vec{b} \, e^{- i \vec{q} \cdot \vec{b} } \, \left[ e^{i \delta (b,s) } -1 \right] \, , 
\ee
where the phase is given by 
\be
\delta(b,s)= \frac{1}{2s} \int \frac{d^{D-2} \vec{q}}{(2 \pi)^{D-2} } e^{i \vec{q} \cdot \vec{b} } A_{\rm tree} (s, - \vec{q}\,^2 ) \, . 
\label{phase}
\ee
It is easy to see that the result is independent on the particular theory (higher derivative or weakly non-local) under the favorable circumstances we are here investigating and remanded above. 

Finally, the Shapiro's time delay is:
\be
\Delta t = 2 \partial_E \delta (E, b) \, . 
\label{ShapiroDT}
\ee
where $E$ is the energy of the probe-particle.

{\em Causality in purely non-local gravity ---} In \cite{yaudong}, the four-graviton scattering amplitude was computed for a large class of local and weakly non-local gravitational theories. It turned out that, for theories that are quadratic in both the Ricci and scalar curvature and lack a term quadratic in the Riemann tensor, 
the tree-level amplitude exactly coincides with the EH one.  From this result, it is already clear that Shapiro's time delay in these theories exactly coincides with the one in general relativity and there is no causality issue independently of the choice of the form factor. 
It deserves to be noticed that there is no causality violation despite we are probing the scale of non-locality.

For completeness we here remind the computation of the time delay in Einstein's gravity. The leading four-graviton amplitude for the purely gravitational Einstein-Hilbert theory in the Regge limit is: $A_t(++,++) =  - 8 \pi G s^2/t$.
Therefore, the phase (\ref{phase}) and the Shapiro's time delay are respectively \cite{CEMZ}:
\be
&& \delta_{\rm g} (b,s) = \frac{\Gamma \left(\frac{D-4}{2} \right)}{\pi^{\frac{D-4}{2} } } \frac{G s}{b^{D-4}} \, 
({\epsilon}_1\cdot {\epsilon}_3)({\epsilon}_2\cdot{\epsilon}_4)
\, , 
\label{phaseEH} \\
&& \Delta t_{\rm g}  =  \frac{\Gamma \left(\frac{D-4}{2} \right)}{\pi^{\frac{D-4}{2} } } \frac{16 E G}{b^{D-4}}\, 
({\epsilon}_1\cdot {\epsilon}_3)({\epsilon}_2\cdot{\epsilon}_4)
\, .
\label{Dtg0}
\ee
where $\epsilon_i$ ($i=1,\dots,4$) are the polarizations of the incoming gravitons 1 and 2 and the outgoing ones 3 and 4.

This result can also be easily understood in terms of the formula in \cite{CEMZ}
\be
&& \hspace{-0.8cm}
\delta_{\rm g} (\vec{b},s) =\frac{\mathcal{A}_3^{I13}(-i\partial_{\vec{b}})\mathcal{A}_3^{I24}(-i\partial_{\vec{b}})}{2s}\int \frac{d^{D-2}\vec{q}}{(2\pi)^{D-2}}\frac{e^{i \vec{q}\cdot\vec{b}}}{\vec{q}^2}\, ,
\nonumber 
 \\
&&=\frac{\Gamma \left(\frac{D-4}{2} \right)}{4\pi^{\frac{D-2}{2} } }\frac{\mathcal{A}_3^{I13}(-i\partial_{\vec{b}})\mathcal{A}_3^{I24}(-i\partial_{\vec{b}})}{2s} \frac1{\vert\vec{b}\vert^{D-4}}\, ,
\ee
where the three-point functions are evaluated on-shell. Notice these functions are in general non-vanishing because the intermediate momentum, which is evaluated on the massless pole, has one imaginary and one real component (we are here assuming that we only have the graviton pole.) Nevertheless, direct inspection of the three-graviton vertices corresponding to terms quadratic in the scalar and Ricci curvatures shows that they are vanishing under such conditions. It is worth noticing that the polarizations of external particles can be viewed as living purely in the transverse space whereas there is only one relevant polarization for the intermediate state, which is perpendicular to the intermediate momentum $\vec{q}$.  So the only relevant three-point function is the one associated with Einstein's gravity, giving (\ref{phaseEH}) and (\ref{Dtg0}). 


{\em Causality violation in Gauss-Bonnet gravity ---} In this section we consider an example of higher derivative theory in $D>4$ dimensions that violates causality. The Lagrangian reads,
\be
\hspace{-0.2cm}
\mathcal{L}
= \frac{2}{\kappa_D^2}
\left[R+ \lambda_{\rm GB}  (R_{\mu\nu\rho\sigma} R^{\mu\nu\rho\sigma} - 4 R_{\mu\nu} R^{\mu\nu} + R^2)  \right]  .
\label{LGB}
\ee
Since only the Einstein-Hilbert term contributes to the propagator, then the scattering amplitude for two gravitons in two gravitons in the Regge limit is the sum of the Einstein-Hilbert's theory one plus a correction due to the Gauss-Bonnet's contribution to the vertex, the result is \cite{Bellazzini}:
\be
&& \label{AGB}
\hspace{-0.5cm} 
A_t
= A_{t {\rm EH} }+ A_{t {\rm GB}} \approx  - \frac{8 \pi G s^2}{t} ({\epsilon}_1\cdot {\epsilon}_3)({\epsilon}_2\cdot{\epsilon}_4) +  \\
&& \hspace{-0.5cm} 
+ \frac{4 \kappa_D^2  \lambda_{\rm GB} s^2}{t} 
\left( k_2^\mu k_4^\nu \epsilon_{2 \nu}^\rho \epsilon_{4 \rho \mu} \epsilon_1 \cdot \epsilon_3 
+ k_1^\mu k_3^\nu \epsilon_{1 \nu}^\rho \epsilon_{3 \rho \mu} \epsilon_2 \cdot \epsilon_4  
\right) .
\nonumber 
\ee
where the momenta of the four gravitons $k_1$, $k_2$, $k_3$, $k_4$ are all incoming, i.e. $\sum_{i=1}^4k_i=0$, and $\epsilon_i$ ($i=1,\dots,4$) are the polarizations of the gravitons. 
We now rewrite the Gauss-Bonnet contribution to the amplitude in terms of the momentum transfer $q = -(k_1+k_3)$ ($t = - q^2$) that, for real kinematics, vanishes in the forward limit. At the leading order in $q$ the amplitudes reads: 
\begin{widetext}
\be
&& 
A_{t {\rm GB}}  \approx  -  \frac{4 \kappa_D^2  \lambda_{\rm GB} s^2    q^\mu q^\nu}{q^2} 
\left(\epsilon_{2 \mu}^\rho \epsilon_{2 \rho \nu} \epsilon_1 \cdot \epsilon_1 
+ \epsilon_{1 \mu}^\rho \epsilon_{1 \rho \nu} \epsilon_2 \cdot \epsilon_2  
\right)      , \nonumber\\
&&
\ee
We notice the following relation between the EH and the GB amplitudes,
\be
A_{t {\rm GB} } (-q^2) = - 4 \lambda_{\rm GB} 
\left(\epsilon_{2 \mu}^\rho \epsilon_{2 \rho \nu} \epsilon_1 \cdot \epsilon_1 
+ \epsilon_{1 \mu}^\rho \epsilon_{1 \rho \nu} \epsilon_2 \cdot \epsilon_2
\right) 
q^\mu q^\nu 
\frac{A_{t {\rm EH}  }(-q^2)}{(\epsilon_1 \cdot \epsilon_{1})  (\epsilon_2 \cdot \epsilon_2)} \, .
\ee
Therefore, we can compiute the phase (\ref{phase}) taking the second derivative of (\ref{phaseEH}) respect to $\vec{b}$,
\be
&& 
\delta_{\rm GB}(b, s) = 4 \lambda_{\rm GB} \left(\epsilon_{2 \mu}^\rho \epsilon_{2 \rho \nu} \epsilon_1 \cdot \epsilon_1 
+ \epsilon_{1 \mu}^\rho \epsilon_{1 \rho \nu} \epsilon_2 \cdot \epsilon_2
\right)  
\partial_{b_i} \partial_{b_j} \frac{\Gamma \left(\frac{D-4}{2} \right)}{\pi^{\frac{D-4}{2} } } \frac{G s}{b^{D-4}} \nonumber \\
&& \hspace{1.45cm}
= - 4 \lambda_{\rm GB} \frac{\Gamma \left(\frac{D-4}{2} \right)}{\pi^{\frac{D-4}{2} } } 
\frac{G s}{b^{D-2}} 
 \left( (\epsilon_1 \cdot \epsilon_{1})  (\epsilon_2 \cdot \epsilon_2) - (D-2)  (\epsilon_2 \cdot \epsilon_2) (n \cdot \epsilon_1) \right) \, , 
 \label{phaseGB}
\ee
where $\vec{n} \equiv \vec{b}/b$. The total contribution to the phase is given by the sum of (\ref{phaseEH}) and (\ref{phaseGB}), namely $\delta_{\rm g-GB} (b,s) = \delta_{\rm g} (b,s) + \delta_{\rm GB} (b,s)$. Finally, the Shapiro's time delay is:
\be
\Delta t_{\rm g-GB} = \frac{\Gamma \left(\frac{D-4}{2} \right)}{\pi^{\frac{D-4}{2} } }  \frac{16 E G}{b^{D-4}}
(\epsilon_1 \cdot \epsilon_{1}) (\epsilon_2 \cdot \epsilon_2)
\left[ 1+ 
\frac{4 \lambda_{\rm GB} (D-2) (D-4)}{b^2} \left( \frac{(n \cdot \epsilon_1)^2}{\epsilon_1 \cdot \epsilon_{1}} - \frac{1}{D-2} \right) \right] \, .
\label{Tgb}
\ee
\end{widetext}
We can see that if the impact factor $b$ becomes small, $b<\lambda_{\rm GB}$, the second term in (\ref{Tgb}) can be bigger then the first one, depending on the sign of $\lambda_{\rm GB}$ and the polarizations. Therefore, we can have a time advance and causality is violated.

{\em Causality in nonlocal gravity coupled to local scalar matter ---} Now we consider ordinary matter described by a two-derivatives theory coupled to nonlocal gravity. Once more we study only a very special class of gravitational theories that are weakly nonlocal, which means they have properties very similar to local two-derivatives theories, namely (\ref{theory}) with form factors (\ref{Hkuzmin}) and (\ref{HTombo}). 
The full action consists of (\ref{theory}) and the minimally coupled ordinary two-derivatives scalar matter,
\be
\hspace{-0.3cm}
S = S_{\rm g} + \int \! d^Dx \sqrt{-g} \left( - \frac{1}{2} g^{\mu\nu} \partial_\mu \phi \partial_\nu \phi - \frac{1}{2} m^2 \phi^2 \right)  .
\label{scalar}
\ee
The graviton propagator is modified by the form factor, but without introducing any extra pole besides the graviton. Therefore, in momentum space and ignoring the gauge dependent terms the propagator reads
\be
G(k) = \frac{e^{-H(\sigma k^2) }}{i (k^2 - i \epsilon) } \left( P^{(2)} - \frac{1}{D-2} P^{(0)} \right)  \, ,
\label{NLP}
\ee
where $P^{(2)}$ and $P^{(0)}$ are the usual spin two and spin zero projector operators \cite{modesto}. 

The tree-level gravitational scattering amplitude for $2$-scalars in $2$-scalars can be easily obtained from the one in local gravity and the result is (we here assume $m=0$):
\be
&& A_s = - 8 \pi G \frac{u t}{s} e^{- H(s)} \, , \quad 
A_t = - 8 \pi G \frac{s (s+ t) }{t} e^{- H(t)} \, , \quad \nonumber 
\\
&& A_u= - 8 \pi G \frac{s t}{u} e^{- H(u)} \, . 
\ee
Notice that the amplitudes are soft in the ultraviolet regime. Therefore, the unitarity bound is satisfied at tree-level. However, in the Regge limit $t \ll s$, which is what we use in the eikonal approximation, the leading contribution comes from the amplitude in the $t$-channel, namely 
\be
A_t(s,t) \approx - 8 \pi G \frac{s^2}{t} e^{- H(t)} \, . \label{matterT}\quad 
\ee
For the sake of simplicity and in order to avoid infrared divergences we here assume $D>4$. However, the result can be easily generalized to a four-dimensional spacetime. 
 For the amplitude (\ref{matterT}) we can now compute the phase (\ref{phase}) in $D=5$, 
 \be
&& \hspace{-1.0cm} 
\delta(b,s)= \frac{1}{2s} \int \frac{d^{3} \vec{q}}{(2 \pi)^3 } e^{i \vec{q} \cdot \vec{b} } A_{t} (s, - \vec{q}\,^2 ) \nonumber \\
&& =  4 \pi G s \int \frac{d^{3} \vec{q}}{(2 \pi)^3 } e^{i \vec{q} \cdot \vec{b} } \, 
\frac{e^{- H(-\vec{q}\,^2) }}{\vec{q}^2} \nonumber \\
&& =  \frac{2  G s}{\pi}   \int dq \frac{\sin (b q)}{b q}  \, 
e^{- H(-q^2) } \, ,
\label{phaseMatter}
 \ee
where $q = |\vec{q}|$. 
We now evaluate (\ref{phaseMatter}) for the form factors (\ref{Hkuzmin}),  (\ref{HTombo}), and also 
\be
 e^{- \sigma \Box} \, , 
\label{SFT}
\ee
which has features very similar to (\ref{Hkuzmin}) and can be analytically integrated. 
Moreover, the form factor (\ref{SFT}) emerges naturally in string field theory 
\cite{collective, Calcagni:2009jb,Calcagni:2013eua,Calcagni:2014vxa}.
Notice that if one expands (\ref{SFT}) in Taylor series, the effective higher derivative theory quadratic in the curvature can potentially violate non only unitarity, but also causality \cite{CEMZ}.

\begin{figure}
\begin{center}
\includegraphics[height=7.5cm]{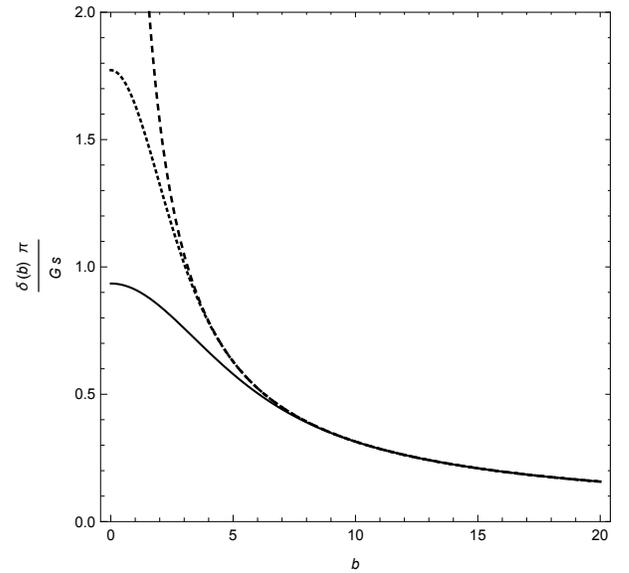}
\end{center}
\caption{From top to bottom the lines represent respectively the following Shapiro's delays: $\Delta t_{\rm EH}$, $\Delta t_{\rm SFT}$, and $\Delta t_{\rm K}$ (for $\alpha =4$). We also assumed $\ell_{\Lambda} =1$.
\label{Fig1}}
\end{figure}
\begin{figure}
\begin{center}
\includegraphics[height=7.5cm]{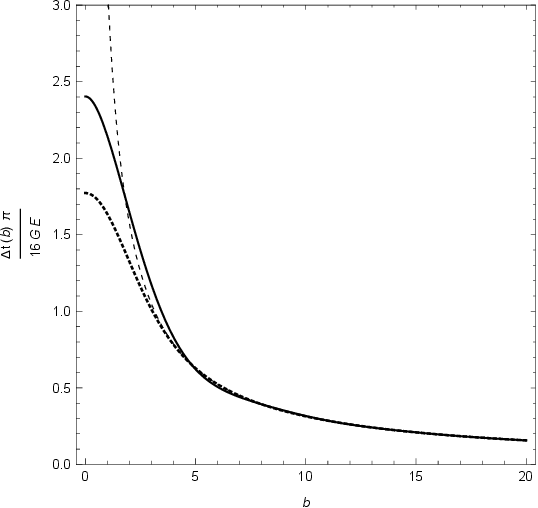}
\end{center}
\caption{From top to bottom the lines represent respectively the following Shapiro's delays: $\Delta t_{\rm EH}$, 
$\Delta t_{\rm T}$ (for $\gamma =3$), and $\Delta t_{\rm SFT}$. We also assumed $\ell_{\Lambda} =1$. 
}
 \label{Fig2}
%
\end{figure}

For the form factor (\ref{SFT}) the phase (\ref{phaseMatter}) after integration gives
\be
\delta (b, s)_{\rm SFT} = G s \, \frac{ {\rm Erf}(b/2\ell_\Lambda)}{b} \, ,
\label{deltaSFT}
\ee
which reduces to the one in Einstein's theory for $b\gg \ell_\Lambda$, namely 
\be
\delta (b, s)_{\rm SFT} \,\, \rightarrow \,\, \delta_{\rm EH} = G s/b.
\ee 
Analytical results can also be obtained for the form factors
\be
e^{- ( -\sigma  \Box)^n} \quad {\rm with} \quad n>1 \, .
\label{SFT2}
\ee
For the case $n=2$, the phase $\delta (b,s)$ reads:
\be
&& \hspace{-0.7cm} 
\delta (b, s)_{{\rm for} (\ref{SFT2})} = G s \Big[ 2 \Gamma \left(\frac{5}{4}\right) \,_1F_3\left(\frac{1}{4};\frac{1}{2},\frac{3}{4},\frac{5}{4}; \frac{b^4}{256 \ell_\Lambda^4}\right) \nonumber \\
&& +\frac{1}{48} b^2 \Gamma \left(-\frac{1}{4}\right) \,_1F_3\left(\frac{3}{4};\frac{5}{4},\frac{3}{2},\frac{7}{4};\frac{b^4}{256 \ell_\Lambda^4 }\right) \Big] \, . 
\ee
The Shapiro's delay is obtained employing formula (\ref{ShapiroDT}) for $s = 4 E$, where $E$ is the energy of the probe-scalar particle. 
For the form factor (\ref{SFT}) and for the Einstein's gravity, the delays read 
\be
&& \Delta t_{\rm SFT}  = \frac{16 E G}{\pi} \, \pi \frac{   {\rm Erf} (b/2\ell_{\Lambda}) }{b} \, , \label{DTsft} \\
&& \Delta t_{\rm EH}  =\frac{ 16 E G}{\pi} \,  \frac{ \pi }{b} \label{DTEH} \, .
\ee
The Shapiro's delay $\Delta t_{\rm K}$ for the form factor (\ref{Hkuzmin}) has been obtained integrating numerically (\ref{phaseMatter}). The result is given in Fig.\ref{Fig1} together with $\Delta t_{\rm SFT}$  and  
$\Delta t_{\rm EH}$.  
In Fig.\ref{Fig2} we plot $\Delta t_{\rm T}$ for the form factor (\ref{HTombo}), which has been obtained numerically, together again with $\Delta t_{\rm SFT}$  and $\Delta t_{\rm EH}$. 
Very similar results can be obtained for different values of $\alpha$ and $\gamma$.

From the analytical results as well as from the plots, it is clear that the Shapiro's time delay never becomes negative and the causality condition is satisfied up to and beyond the non locality scale $\ell_\Lambda$.

{\em General causal nonlocal theories ---} 
In this section we shortly remind a field redefinition theorem \cite{AnselmiFRT, Dona}  that allows us to map nonlocal field theories to local ones at tree-level. 
The theorem is perturbative in the field redefinition, but the identification is only valid at tree-level because of the different quantum properties of the two theories. 
%

Let us consider two general weakly nonlocal actions functionals 
$S'(g, \Phi_a)$  and $S(g',\Phi'_a)$, respectively defined in terms of the fields 
$g, \Phi_a$ and $g',\Phi'_a$, where $g$ is the metric and $\Phi_a$ a set of matter or gauge fields and, and such that 
\be
&& 
\hspace{-0.8cm}
S'(g, \Phi_a) = 
 S(g, \Phi_a) + E^g_i(g, \Phi_a) F^g_{i j}(g, \Phi_a)  E^g_j (g, \Phi_a) \nonumber \\
&& \hspace{1cm}
+ E^\Phi_a(g, \Phi_c) F^\Phi_{a b}(g, \Phi_c)  E^\Phi_b (g, \Phi_c) 
 \, ,
 \label{AnselmiC}
\ee
where $F^g$ and $F^\Phi$ can contain derivative operators or weakly nonlocal operators of the covariant 
$\Box$ operator, and 
\be
E^g_i = \frac{\delta S}{\delta g_i}  \, , \quad E^\Phi_a= \frac{\delta S}{\delta \Phi_a} 
\ee
are the EOM of the theory with action $S(g, \Phi_a)$. 
Here we use a compact deWitt notation and with the indices $i$, $a$ on fields we encode all Lorentz, group indices, and the spacetime dependence of the fields. 
Notice that we here mean the metric in the fields' space and not 
the spacetime metric $g_{\mu\nu}$. 
%
The statement of the theorem is that there exists a field redefinition 
\be
&& g_i'  = g_i + \Delta^g_{i j} E^g_j  \, , \qquad \Delta^g_{i j} = \Delta^g_{j i}, \nonumber \\
&& \Phi_a'  = \Phi_a + \Delta^\Phi_{a b} E^\Phi_b \, , \qquad \Delta^\Phi_{a b} = \Delta^\Phi_{b a} \, , 
\label{FR2}
\ee
such that, perturbatively in ${F^g,\Phi}$, but to all orders in powers of $F^{g,\Phi}$, we have the equivalence
\be
S'(g) = 
S(g')  \,.
\label{FR}
\ee
$\Delta^g_{ij}$ ($\Delta^\Phi_{ab}$) could be a weakly nonlocal or quasi-polynomial operator acting linearly on the EOM $E^g_j$ ($E^\Phi_a$), 
and they are defined perturbatively 
in powers of the operators $F^{g, \Phi}$, namely 
\be
\Delta^g_{ij}=F^g_{ij}+\ldots \quad {\rm or} \qquad \Delta^\Phi_{ab}=F^\Phi_{ab}+\ldots \, .
\label{FRcoefficients}
\ee
For the sake of simplicity we can check the claim above at the first order in the Taylor expansion for the functional $S(g', \Phi'_a)$, which reads
\be
&& \hspace{-1.5cm}
S(g', \Phi'_a) = S(g + \Delta g, \Phi + \Delta \Phi ) \nonumber \\
&& \approx S(g) 
+ \frac{\delta S}{\delta g_i}  \delta g_i + \frac{\delta S}{\delta \Phi_a}  \delta \Phi_a 
\nonumber \\
&& = S(g) + E^g_i   \, \delta g_i  + E^\Phi_a   \, \delta \Phi_a  
 \, .
\ee
Therefore, the field redefinitions \eqref{FR2} with the choicen coefficients \eqref{FRcoefficients} are such that 
the equivalence \eqref{FR} is satisfied (note that the argument of the functionals $S'$ and $S$ is now the same.) Hence the two actions $S'(g)$ and $S(g')$ are tree-level equivalent.

A crucial consequence of the theorem just reminded is that all the $n$-points tree-level functions of the theories related by a field redefinition are identical. Therefore, the causality condition given in \cite{CEMZ} comes as a mere consequence of the four-points functions 
computed for the local theory. 

Finally, {\em any nonlocal theory that is tree-level equivalent to a causal local one is causal too. In other words,
in this section we provided an algorithm for constructing higher derivative (even non-local) causal theories.}

We now provide an explicit example of non-local theory involving gravity, one gauge field, and a scalar field that is unitary and finite at quantum level in odd dimension (in particular in $D=5$.) Moreover, the field redefinition theorem ensures the causality condition based on Shapiro's time delay because the $n$-points scattering amplitudes are the same of the tree-level equivalent theory. The Lagrangian reads:
\be
&& \hspace{-0.5cm} 
\mathcal{L} =  \frac{2}{\kappa_D^2} \left[ R + \left( G_{\mu\nu} - \frac{\kappa_D^2}{2} (T^{A}_{\mu\nu} +  T^{\phi}_{\rho \sigma} )
\right) \times 
\right. \label{GT} \\
&& \hspace{0.7cm} 
\left. \times 
F_{\rm g}^{\mu\nu, \rho \sigma} \left( G_{\rho \sigma} - \frac{\kappa_D^2}{2} (T^{A}_{\rho \sigma} +  T^{\phi}_{\rho \sigma} )  \right) \right] \nonumber \\
&&  \hspace{0.7cm} 
- \frac{1}{4} F_{\mu\nu}F^{\mu\nu} +  \nabla_\mu F^{\mu\nu} \, F^{\rm A} \, \nabla_\rho F^{\rho}{}_{\nu}  \nonumber \\
&&  \hspace{0.7cm} 
+ \frac{1}{2} \phi (\Box - m^2) \phi + \phi (\Box - m^2) \,  F^\phi \, (\Box - m^2) \phi \,  ,
\nonumber 
\ee
where the analytic functions of the d'Alembertian operator $F^{\rm g}$, $F^A$, $F^\phi$ and the second rank tensors $T^{A}_{\mu\nu}$, $T^{\phi}_{\mu\nu}$ are defined as follows, 
\be
&& F_{\rm g}^{\mu\nu, \rho \sigma} \equiv \left( g^{\mu \rho} g^{\nu\sigma} - \frac{1}{2} g^{\mu \nu} g^{\rho \sigma} \right) \left( \frac{e^{H_{\rm g} (\Box) } -1}{\Box} \right)  \, , \nonumber \\
&&
  F^{\rm A} \equiv  \frac{1}{2} \left( \frac{e^{H_{A} (\Box) } -1}{\Box} \right)   \, , \nonumber \\
  &&
F^\phi \equiv \frac{1}{2} \left( \frac{e^{H_\phi(\Box - m^2)} -1 }{\Box - m^2} \right) \, , \nonumber \\
&&  T^{A}_{\mu\nu} \equiv F_{\mu\sigma} F^{\sigma}_\nu - \frac{1}{4} F_{\mu\nu}F^{\mu\nu} \,  , \nonumber \\
&& T^{\phi}_{\mu\nu} \equiv  \partial_{\mu} \phi  \partial_\nu \phi 
- \frac{1}{2} g_{\mu\nu} (\partial_\lambda \phi \partial^\lambda \phi + m^2 \phi^2 ) 
\, .
\ee
It is straightforward to prove that the theory above satisfied the field redefinition theorem, namely it is equivalent to Einstein's gravity minimally coupled to electromagnetic field and scalar matter. Therefore, all the tree-level $n$-point functions for the theory (\ref{GT}) are identical to the ones of local gravity couple to the local Maxwell field. 

Supersymmetry is another powerful tool to couple nonlocal gravity to matter in a way consistent with unitarity, finiteness, but also causality. Indeed, in \cite{Giaccari} we constructed the $N=1$ nonlocal supergravity that has the same tree-level scattering amplitudes as the local one. Therefore, on the basis of the theorem reviewed in this section the theory is causal.

{\em Local Lee-Wick quantum gravity ---} Recently a local higher derivative theory has been proposed as a good candidate for a UV completion of the Einstein-Hilbert theory \cite{shapiromodesto, LWqg}. 
The theory has no real ghosts in the spectrum, but allows for complex conjugate ghosts. It turns out that it is unitary at tree-level \cite{shapiromodesto, LWqg} and also at any perturbative order in the loop expansion \cite{Piva1,Piva2,Piva3}. Moreover, it is super-renormalizable or finite at quantum level \cite{shapiromodesto,LWqg}. Notice that the $S$-matrix is unitary in the subspace of real states as a consequence of the energy conservation and on the basis of the empirical evidence that complex energy is not realized in nature. 

The minimal theory in which only the graviton propagates and a pair of complex conjugate ghosts reads:
\be
S_{\rm g}  = \frac{2}{\kappa^{2}_D}  \int  d^D x \sqrt{-g} \left[ R + \sigma^2 G_{\mu\nu} \Box R^{\mu\nu} + V(\mathcal{R}) \right]  .
\label{LWqg}
\ee
The propagator of the theory (\ref{LWqg}) is very similar to (\ref{NLP}), but shows up two complex conjugate poles, 
\be
\hspace{-0.2cm}
G(k) = \frac{1}{i (k^2 - i \epsilon) (1+\sigma^2 k^4)} \left( P^{(2)} - \frac{1}{D-2} P^{(0)} \right)  .
\ee
If the potential $V(\mathcal{R})$ is at least quadratic in the Ricci tensor the tree-level amplitudes coincide with 
those in Einstein gravity (this is a consequence of the theorem in the previous section) and causality is not violated. 
When the Lee-Wick gravity (\ref{LWqg}) is coupled to the scalar matter (\ref{scalar}) 
the tree-level gravitational scattering amplitude for $2$-scalars in $2$-scalars is obtained from the amplitude in the $t$-channel 
\be
A_t = - 8 \pi G \frac{s (s+ t) }{t (1+ \sigma^4 t^2)}  \approx - 8 \pi G \frac{s^2 }{t (1+ \sigma^4 t^2)}\, . 
\label{ALW}
\ee
Notice that $t\ll s$, but $t$ can be larger then $\Lambda^2$. Replacing the amplitude (\ref{ALW}) in (\ref{phaseMatter}) we get
\be
\delta (b, s)_{\rm SFT} = G s \, \frac{ 1 - e^{-\frac{b}{2\ell_\Lambda} } \cos (\frac{b}{\sqrt{2} \ell_\Lambda})  }{b} \, .
\label{deltaLW}
\ee
The phase (\ref{deltaLW}) is always positive and the plot is very similar to the one of nonlocal gravity. Therefore, the Shapiro's time delay is also positive and there is no causality violation.

{\em Nonlocal gravity in Weyl basis ---} Finally, we would like to present a theory that can potentially violate causality. The Lagrangian reads 
\cite{yaudong,ModestoConformal, Koshelev:2016xqb, Koshelev:2017tvv, Biswas:2013cha}:
\be
&& \hspace{-1cm}
 \mathcal{L}_{W} = \frac{2}{\kappa_D^2} \left( R + C_{\mu\nu\rho\sigma} \gamma_{\rm C}(\Box) C^{\mu\nu\rho\sigma} + R \gamma_{\rm R} (\Box) R \right) \label{WeylT} \, , \nonumber \\
&& \hspace{-1cm}
\gamma_{\rm C} =   \frac{D-2}{4(D-3)}  \frac{e^{H_2} -1}{\Box} \, ,  \,\,
\gamma_{\rm S} = - \frac{D-2}{4(D-1)}  \frac{e^{H_0} -1}{\Box}  .
\ee
The theory (\ref{WeylT}) violates causality for $H_2 = H_0 = H_{\rm K}$ or $H_2 = H_0 = \sigma \Box$. Indeed, expanding $\gamma_{\rm C}$ in Taylor series 
we get also a Riemann square operator, $R_{\mu\nu\rho\sigma} R^{\mu\nu\rho\sigma}$, that gives the same causality violation as computed in (\ref{Tgb}). On the other hand, the same theory with entire functions $H_2 = H_0 = H_{\rm T}$ does not give a Shapiro's time advanced as shown in \cite{CEMZ}. 

{\em Conclusions ---}
We proved that weakly nonlocal pure gravity, a theory proposed as an ultraviolet completion of Einstein's gravity \cite{modestoLeslaw}, is causal by the means of Shapiro's time delay. Indeed, we do not detect any time advancement (which could allow for the construction of a time machine.) 
Moreover, ordinary matter described by a two derivative theory is also compatible with causality when coupled to nonlocal gravity. Finally, we provided a recipe based on a field redefinition theorem (FRT) to construct a causal general nonlocal theory for matter coupled to gravity. Consistently with the FRT and on the basis of explicit computations we proved that Shapiro's causality is fulfilled in Lee-Wick gravity too.


\begin{acknowledgments}
The research of S.G. was supported by the Israel Science Foundation (ISF), grant No. 244/17. S.G. is also thankful to the University of Zagreb and the Croatian Science Foundation for support under the project No. 8946 during the first part of this research.
\end{acknowledgments}



\begin{thebibliography}{99}

\bibitem{CEMZ} 
  X.~O.~Camanho, J.~D.~Edelstein, J.~Maldacena and A.~Zhiboedov,
  ``Causality Constraints on Corrections to the Graviton Three-Point Coupling,''
  JHEP {\bf 1602}, 020 (2016)
  doi:10.1007/JHEP02(2016)020
  [arXiv:1407.5597 [hep-th]].

  


  
\bibitem{kuzmin} 
  Y.~V.~Kuz'min,
  ``The Convergent Nonlocal Gravitation. (in Russian),''
  Sov.\ J.\ Nucl.\ Phys.\  {\bf 50}, 1011 (1989)
  [Yad.\ Fiz.\  {\bf 50}, 1630 (1989)].



\bibitem{tomboulis}
  E.~T.~Tomboulis,
  ``Renormalization and unitarity in higher derivative and nonlocal gravity theories,''
  Mod.\ Phys.\ Lett.\ A {\bf 30}, 1540005 (2015). E.~T.~Tomboulis, hep-th/9702146. 
  
\bibitem{modesto}
  L.~Modesto,
  ``Super-renormalizable Quantum Gravity,''
  Phys. \ Rev. \ D {\bf 86}, 044005 (2012)
  [arXiv:1107.2403 [hep-th]].
  
   \bibitem{modestoLeslaw}
  L.~Modesto and L.~Rachwal,
  ``Super-renormalizable and finite gravitational theories,''
  Nucl.\ Phys.\ B {\bf 889}, 228 (2014) 
  [arXiv:1407.8036 [hep-th]].


%
%

  

  
 

\bibitem{Erice} 
  S.~B.~Giddings,
  ``The gravitational S-matrix: Erice lectures,''
  Subnucl.\ Ser.\  {\bf 48}, 93 (2013)
  [arXiv:1105.2036 [hep-th]].
 

\bibitem{GiddingsPorto} 
  S.~B.~Giddings and R.~A.~Porto,
  ``The Gravitational S-matrix,''
  Phys.\ Rev.\ D {\bf 81}, 025002 (2010)
  [arXiv:0908.0004 [hep-th]].


  
 
\bibitem{Kabat} 
  D.~N.~Kabat and M.~Ortiz,
  ``Eikonal quantum gravity and Planckian scattering,''
  Nucl.\ Phys.\ B {\bf 388}, 570 (1992)
  [hep-th/9203082].
 
 
\bibitem{Ciafaloni} 
  M.~Ciafaloni and D.~Colferai,
  ``Rescattering corrections and self-consistent metric in Planckian scattering,''
  JHEP {\bf 1410}, 85 (2014)
  [arXiv:1406.6540 [hep-th]].
 
 
 
\bibitem{Bellazzini} 
  B.~Bellazzini, C.~Cheung and G.~N.~Remmen,
  ``Quantum Gravity Constraints from Unitarity and Analyticity,''
  Phys.\ Rev.\ D {\bf 93}, no. 6, 064076 (2016)
  [arXiv:1509.00851 [hep-th]].
 
 
 \bibitem{yaudong} 
  Y.~D.~Li, L.~Modesto and L.~Rachwal,
  ``Exact solutions and spacetime singularities in nonlocal gravity,''
  JHEP {\bf 1512}, 173 (2015)
  [arXiv:1506.08619 [hep-th]].    
  


 
\bibitem{Giaccari} 
  S.~Giaccari and L.~Modesto,
  ``Nonlocal supergravity,''
  Phys.\ Rev.\ D {\bf 96}, no. 6, 066021 (2017)
  [arXiv:1605.03906 [hep-th]].
 


\bibitem{shapiromodesto} 
  L.~Modesto and I.~L.~Shapiro,
  ``Superrenormalizable quantum gravity with complex ghosts,''
  Phys.\ Lett.\ B {\bf 755}, 279 (2016)
  [arXiv:1512.07600 [hep-th]]. 

\bibitem{LWqg} 
  L.~Modesto,
  ``Super-renormalizable or finite Lee-Wick quantum gravity,''
  Nucl.\ Phys.\ B {\bf 909}, 584 (2016)
  [arXiv:1602.02421 [hep-th]].  



\bibitem{collective} 
  V.~A.~Kostelecky and S.~Samuel,
  ``Collective Physics in the Closed Bosonic String,''
  Phys.\ Rev.\ D {\bf 42}, 1289 (1990).


\bibitem{Calcagni:2009jb} 
  G.~Calcagni and G.~Nardelli,
  ``String theory as a diffusing system,''
  JHEP {\bf 1002}, 093 (2010)
  [arXiv:0910.2160 [hep-th]].

\bibitem{Calcagni:2013eua} 
  G.~Calcagni and L.~Modesto,
  ``Nonlocality in string theory,''
  J.\ Phys.\ A {\bf 47}, no. 35, 355402 (2014)
  [arXiv:1310.4957 [hep-th]].
  
\bibitem{Calcagni:2014vxa} 
  G.~Calcagni and L.~Modesto,
  ``Nonlocal quantum gravity and M-theory,''
  Phys.\ Rev.\ D {\bf 91}, no. 12, 124059 (2015)
  [arXiv:1404.2137 [hep-th]].

  
  
  

\bibitem{AnselmiFRT} 
  D.~Anselmi and M.~Halat,
  ``Renormalizable acausal theories of classical gravity coupled with interacting quantum fields,''
  Class.\ Quant.\ Grav.\  {\bf 24}, 1927 (2007)
  [hep-th/0611131].
  
\bibitem{Dona} 
  P.~Donà, S.~Giaccari, L.~Modesto, L.~Rachwal and Y.~Zhu,
  ``Scattering amplitudes in super-renormalizable gravity,''
  JHEP {\bf 1508}, 038 (2015)
  [arXiv:1506.04589 [hep-th]].
  
  
    
  
\bibitem{Piva1} 
  D.~Anselmi and M.~Piva,
  ``A new formulation of Lee-Wick quantum field theory,''
  JHEP {\bf 1706}, 066 (2017)
  [arXiv:1703.04584 [hep-th]].

  
\bibitem{Piva2} 
  D.~Anselmi and M.~Piva,
  ``Perturbative unitarity of Lee-Wick quantum field theory,''
  Phys.\ Rev.\ D {\bf 96}, no. 4, 045009 (2017)
  [arXiv:1703.05563 [hep-th]].

\bibitem{Piva3} 
  D.~Anselmi,
  ``Fakeons And Lee-Wick Models,''
  JHEP {\bf 1802}, 141 (2018)
  [arXiv:1801.00915 [hep-th]].

  
  
  
\bibitem{ModestoConformal} 
  L.~Modesto and L.~Rachwal,
  ``Finite Conformal Quantum Gravity and Nonsingular Spacetimes,''
  arXiv:1605.04173 [hep-th].
  
\bibitem{Koshelev:2016xqb} 
  A.~S.~Koshelev, L.~Modesto, L.~Rachwal and A.~A.~Starobinsky,
  ``Occurrence of exact $R^2$ inflation in non-local UV-complete gravity,''
  JHEP {\bf 1611}, 067 (2016)
  [arXiv:1604.03127 [hep-th]].

\bibitem{Koshelev:2017tvv} 
  A.~S.~Koshelev, K.~Sravan Kumar and A.~A.~Starobinsky,
  ``$R^2$ inflation to probe non-perturbative quantum gravity,''
  arXiv:1711.08864 [hep-th].

\bibitem{Biswas:2013cha} 
  T.~Biswas, A.~Conroy, A.~S.~Koshelev and A.~Mazumdar,
  ``Generalized ghost-free quadratic curvature gravity,''
  Class.\ Quant.\ Grav.\  {\bf 31}, 015022 (2014)
  Erratum: [Class.\ Quant.\ Grav.\  {\bf 31}, 159501 (2014)]
  [arXiv:1308.2319 [hep-th]].


  
  

  
\end{thebibliography}
\end{document}